\documentclass[12pt,fleqn]{article}
\usepackage{graphics,graphicx,cite}

\usepackage{amssymb}

\def\onecol{\onecolumn \mathindent 2em}

\def\noi{\noindent}

\makeatletter
\renewcommand{\section}{\@startsection{section}{1}{0pt}%
        {-3.5ex plus -1ex minus -.2ex}{2.3ex plus .2ex}%
        {\large\bf\protect\raggedright}}

\renewcommand{\subsection}{\@startsection{subsection}{2}{0pt}%
        {-3ex plus -1ex minus -.2ex}{1.4ex plus .2ex}%
        {\normalsize\bf\protect\raggedright}}

\renewcommand{\thesubsubsection}%
        {\arabic{section}.\arabic{subsection}.\arabic{subsubsection}.}

\renewcommand{\@oddhead}{\raisebox{0pt}[\headheight][0pt]{%
   \vbox{\hbox to\textwidth{\rightmark \hfil \rm \thepage \strut}\hrule}}}
\renewcommand{\@evenhead}{\raisebox{0pt}[\headheight][0pt]{%
   \vbox{\hbox to\textwidth{\thepage \hfil \leftmark \strut}\hrule}}}

\makeatother

\def\prepno#1#2
    {\thispagestyle{empty}
    \noi    \unitlength=1mm
    	\begin{picture}(178,10)
            \put(177,15){\llap{\bf #1}}
            \put(177,10){\llap{\small\rm #2}}
        \end{picture}
          }             

\newcommand{\Title}[1]{\noi {\uppercase{\Large #1}} \\}
\newcommand{\Author}[2]{\noi{\large\bf #1}\\[2ex]\noindent{\it #2}\\}

\newcommand{\Abstract}[1]{\vskip 2mm \begin{center}
        \parbox{16.4cm}{\small\noi #1} \end{center}\medskip}

\newcommand{\foom}[1]{\protect\footnotemark[#1]}

\newcommand{\email}[2]{\footnotetext[#1]{e-mail: #2}
		\addtocounter{footnote}{1}}

\newcommand{\sect}[1]{Sec.\,#1}
\def\nq{\hspace*{-1em}}
\def\nqq{\hspace*{-2em}}
\def\nhq{\hspace*{-0.5em}}

\def\cm{\hspace*{1cm}}

\def\Jl#1#2{{\it #1\/} {\bf #2},\ }

\def\ApJ#1 {\Jl{Astroph. J.}{#1}}
\def\CQG#1 {\Jl{Class. Quantum Grav.}{#1}}
\def\DAN#1 {\Jl{Dokl. AN SSSR}{#1}}
\def\GC#1 {\Jl{Grav. \& Cosmol.}{#1}}
\def\GRG#1 {\Jl{Gen. Rel. Grav.}{#1}}
\def\JETF#1 {\Jl{Zh. Eksp. Teor. Fiz.}{#1}}
\def\JETP#1 {\Jl{Sov. Phys. JETP}{#1}}
\def\JHEP#1 {\Jl{JHEP}{#1}}
\def\JMP#1 {\Jl{J. Math. Phys.}{#1}}
\def\NPB#1 {\Jl{Nucl. Phys.}{B\ #1}}
\def\NP#1 {\Jl{Nucl. Phys.}{#1}}
\def\PLA#1 {\Jl{Phys. Lett.}{#1A}}
\def\PLB#1 {\Jl{Phys. Lett.}{#1B}}
\def\PRD#1 {\Jl{Phys. Rev.}{D\ #1}}
\def\PRL#1 {\Jl{Phys. Rev. Lett.}{#1}}

\newcommand{\eqsection}{\makeatletter
	\@addtoreset{equation}{section}
	\renewcommand{\theequation}{\arabic{section}.\arabic{equation}}
	\makeatother}
\def\al{&\nhq}
\def\lal{&&\nqq {}}
\def\eq{Eq.\,}
\def\eqs{Eqs.\,}
\def\beq{\begin{equation}}
\def\eeq{\end{equation}}
\def\bear{\begin{eqnarray}}
\def\bearr{\begin{eqnarray} \lal}
\def\ear{\end{eqnarray}}
\def\earn{\nonumber \end{eqnarray}}

\def\nnn{\nonumber\\ \lal }

\def\eql{\al =\al}

\def\tst{\textstyle}

\def\fract#1#2{{\tst\frac{#1}{#2}}}

\def\half{{\fract{1}{2}}}

\def\d{\partial}

\def\sign{\mathop{\rm sign}\nolimits}

\def\const{{\rm const}}
\def\eps{\varepsilon}

\def\og{\overline{g}}

\def\oR{\overline{R}}

\def\mn{_{\mu\nu}}
\def\MN{^{\mu\nu}}
\def\mN{_\mu^\nu}

\def\ep{\epsilon}

\def\wh{wormhole}
\def\whs{wormholes}
\def\bh{black hole}
\def\bhs{black holes}

\def\ssph{static, spherically symmetric}
\def\asflat{asymptotically flat}

\def\bw{brane world}

\def\BD{Brans-Dicke}
\def\GR{general relativity}

\def\KS{Kantowski-Sachs}
\def\RN{Reiss\-ner-Nord\-str\"om}
\def\Sch{Schwarz\-schild}

\begin{document}
\onecol
\prepno{gr-qc/0611022}{}

\Title {Regular black holes and black universes}

\Author{K.A. Bronnikov\foom 1 and V.N. Melnikov\foom 2}
       {\small Centre for Gravitation and Fundamental Metrology,
 	 VNIIMS, 46 Ozyornaya St., Moscow 119361, Russia;\\
   Institute of Gravitation and Cosmology, PFUR,
        6 Miklukho-Maklaya St., Moscow 117198, Russia}
   \email 1 {kb20@yandex.ru}
   \email 2 {melnikov@phys.msu.ru}

\Author{Heinz Dehnen\foom 3}
  {\small Fachbereich Physik, Universit\"at Konstanz, Postfach M 677 78457
   Konstanz, Germany}
   \email 3 {Heinz.Dehnen@uni-konstanz.de}

\Abstract
  {We give a comparative description of different types of regular \ssph\
  \bhs\ (BHs) and discuss in more detail their particular type, which we
  suggest to call black universes. The latter have a Schwarzschild-like
  causal structure, but inside the horizon there is an expanding \KS\
  universe and a de Sitter infinity instead of a singularity. Thus a
  hypothetic BH explorer gets a chance to survive. Solutions of this kind
  are naturally obtained if one considers \ssph\ distributions of various
  (but not all) kinds of phantom matter whose existence is favoured by
  cosmological observations. It also looks possible that our Universe has
  originated from phantom-dominated collapse in another universe and
  underwent isotropization after crossing the horizon.  An explicit example
  of a black-universe solution with positive Schwarzschild mass is
  discussed.}


\section{Introduction\label{s1}}

  One of the long-standing problems in \bh\ (BH) physics is the existence of
  curvature singularities beyond the event horizon in the BH solutions
  obtained under the simplest and the most natural physical conditions
  (the \Sch, \RN, Kerr and Kerr-Newman solutions of \GR\ and their
  counterparts in many alternative theories of gravity). Singularities are
  places where general relativity (or another classical theory of gravity)
  does not work. Therefore, a full understanding of BH physics requires
  avoidance of singularities or/and modification of the corresponding
  classical theory and addressing quantum effects. There have been numerous
  attempts on this trend, many of them suggesting that a singularity
  inside the event horizon should be replaced with a kind of regular core.

  In this paper, we discuss the possible geometry of classical nonsingular
  BHs, restricting ourselves to \asflat\ \ssph\ configurations. In \sect
  \ref{s2} we enumerate and briefly describe different known types of
  regular BHs. The rest of the paper is devoted to their particular type
  which we suggest to call {\it black universes,\/} namely, those in which a
  possible BH explorer, after crossing the event horizon, gets into an
  expanding universe \cite{pha1}. Such objects are naturally obtained if one
  considers local concentrations of dark energy in the form of phantom
  matter, favoured by modern cosmological observations \cite{pha-obs}.
  Among theoretical reasons for considering phantom matter one may mention
  natural appearance of phantom fields in some models of string theory
  \cite{sen}, supergravities \cite{sugr} and theories in more than 11
  dimensions like F-theory \cite{khvie}. To avoid the obvious quantum
  instability, a phantom scalar may perhaps be regarded as an effective
  field description following from an underlying theory with positive
  energies \cite{no03trod}. Curiously, a classical massless phantom field
  even shows a more stable behaviour than its usual counterpart
  \cite{cold,picon}.

  In \sect \ref{s3} we demonstrate the existence of black-universe
  solutions in \GR\ with minimally coupled phantom fields, formulate the
  requirements to the field potential needed for obtaining them and present
  a specific example. In \sect \ref{s4} we obtain similar requirements in
  a more general framework, general k-essence, and show that many (though
  not all) forms of phantom matter discussed in the recent cosmological
  literature can also lead to black universes. \sect \ref{s5} is a
  conclusion.

\section{Regular \bh\ geometries\label{s2}}

    We begin with the general \ssph\ metric
\beq                                                            \label{ds}
    ds^2 = A(\rho) dt^2 - \frac{d\rho^2}{A(\rho)} - r^2(\rho)d\Omega^2,
\eeq
    where $d\Omega^2 = d\theta^2 + \sin^2\theta d\varphi^2$ is the metric on
    a unit sphere. Our sign conventions are as follows: the metric signature
    $({+}\ {-}\ {-}\ {-})$ and the curvature tensor
    $R^{\sigma}{}_{\mu\rho\nu} = \d_\nu\Gamma^{\sigma}_{\mu\rho}-\ldots$, so
    that, e.g., the Ricci scalar $R = R\MN{}\mn > 0$ for de Sitter
    space-time.

    The metric (\ref{ds}) is written in terms of the ``quasiglobal''
    coordinate $\rho$, which is particularly convenient for dealing with
    Killing horizons where it behaves in the same way as the manifestly
    well-behaved Kruskal-like null coordinates. For this reason, in terms of
    $\rho$, one may consider regions on both sides of such a horizon
    remaining in a formally static framework.

    To discuss BHs in the comparatively simple case of static spherical
    symmetry, we may leave aside more general and more rigorous definitions
    of horizons and \bhs\ (see, e.g., \cite{wald}) and rely on the following
    working definition. A black hole is a space-time containing (i) a static
    region which may be regarded external (e.g., contains a flat
    asymptotic), (ii) another region invisible to an observer at rest
    residing in the static region, and (iii) a Killing horizon of nonzero
    area that separates the two regions and admits an analytical extension
    of the metric from one region to another.

    The two functions, $A(\rho)$ (often called the redshift function) and
    $r(\rho)$ (the area function, equal to the radius of a coordinate sphere
    at given $\rho$) entirely determine the geometry under consideration.
    Asymptotic flatness is described, without loss of generality, by $A \to
    1$ and $r\approx \rho$ as $\rho \to \infty$. A centre $\rho=\rho_c$ (if
    any) corresponds to $r=0$ while Killing horizons (if any) are described
    by zeros of the function $A(\rho)$. Killing horizons, at which the
    timelike Killing vector becomes null, divide the whole space-time
    manifold into static (R) regions, in which $A > 0$, and nonstatic,
    homogeneous (T) regions whose geometry is that of a \KS\ anisotropic
    cosmological model.  The number, order and disposition of such zeros
    determine the global causal structure of space-time.

    The following types of geometries of regular, 4-dimensional,
    \asflat, \ssph\ BHs are known in the literature:

\begin{enumerate}
\item
	BHs with a regular centre ($r\approx \const\cdot \rho-\rho_c$,
	$A \to A_c >0$, $A(dr/d\rho)^2 \approx 1 + O(r^2)$ as
	$\rho\to\rho_c$). Since a regular centre can only be located in an
	R region, such a BH must have at least two simple horizons or one
	double horizon, and its causal structure is then represented by the
	same Carter-Penrose diagram as that of the non-extreme or extreme
	Reissner-Nordstr\"om BH (diagrams 1b and 1c in Fig.\,1),
	respectively. A larger number of horizons is not excluded,
	leading to more complex causal structures.
\item
	BHs without a centre, having second-order horizons of infinite area
	(so-called cold BHs because such horizons are always characterized
	by zero Hawking temperature) \cite{cold,cold-q}. They may have
	different causal structures. In one case, the Carter-Penrose diagram
	coincides with that of Kerr's extreme BH, consisting of an infinite
	tower of R regions but certainly without a ring singularity which
	is present in Kerr's solution (plots 2a and diagram 2b in Fig.\,1).
	In another case, there are only four R regions (plots 2c and diagram
	2d).
\item
	BHs whose causal structure coincides with that of a non-extreme Kerr
	BH, again without a singular ring \cite{casadio,bbh1} (diagram 3b).
\item
	Regular BHs with a Schwarzschild-like causal structure \cite{pha1}
        (diagram 4b) but with cosmological expansion instead of a singular
	centre.
\end{enumerate}

\begin{figure}
\centering
\unitlength=1.00mm
\special{em:linewidth 0.4pt}
\linethickness{0.4pt}
\begin{picture}(156.00,159.00)
\put(10.00,126.00){\vector(0,1){30.00}}
\put(8.00,135.00){\vector(1,0){32.00}}
\put(25.00,84.00){\vector(0,1){37.00}}
\put(8.00,103.00){\vector(1,0){35.00}}
\put(106.00,83.00){\vector(0,1){37.00}}
\put(89.00,102.00){\vector(1,0){35.00}}
\put(25.00,41.00){\vector(0,1){37.00}}
\put(8.00,60.00){\vector(1,0){35.00}}
\put(25.00,-1.00){\vector(0,1){37.00}}
\put(8.00,18.00){\vector(1,0){35.00}}
\bezier{140}(10.00,135.00)(15.00,145.00)(36.00,157.00)
\bezier{72}(10.00,145.00)(15.00,146.00)(21.00,135.00)
\bezier{84}(21.00,135.00)(25.00,126.00)(32.00,135.00)
\bezier{84}(32.00,135.00)(40.00,146.00)(47.00,147.00)
\bezier{64}(10.00,148.00)(14.00,149.00)(21.00,140.00)
\bezier{96}(21.00,140.00)(26.00,130.00)(33.00,141.00)
\bezier{48}(33.00,141.00)(37.00,146.00)(43.00,147.00)
\bezier{160}(9.00,119.00)(22.00,103.00)(23.00,122.00)
\bezier{176}(27.00,122.00)(30.00,102.00)(43.00,122.00)
\bezier{52}(21.00,106.00)(25.00,101.00)(30.00,106.00)
\bezier{160}(90.00,116.00)(103.00,100.00)(104.00,119.00)
\bezier{68}(86.00,110.00)(95.00,109.00)(101.00,104.00)
\bezier{20}(101.00,104.00)(103.00,103.00)(106.00,102.00)
\bezier{24}(106.00,102.00)(109.00,102.00)(111.00,100.00)
\bezier{76}(111.00,100.00)(117.00,94.00)(127.00,93.00)
\bezier{172}(108.00,83.00)(110.00,103.00)(126.00,86.00)
\bezier{64}(6.00,70.00)(13.00,67.00)(18.00,60.00)
\bezier{108}(18.00,60.00)(25.00,49.00)(33.00,60.00)
\bezier{76}(33.00,60.00)(41.00,70.00)(47.00,70.00)
\bezier{228}(48.00,28.00)(22.00,27.00)(6.00,0.00)
\put(75.00,159.00){\line(-1,-1){7.00}}
\put(68.00,152.00){\line(1,-1){7.00}}
\put(75.00,145.00){\line(0,-1){14.00}}
\put(75.00,131.00){\line(-1,-1){5.00}}
\put(75.00,159.00){\line(1,-1){7.00}}
\put(82.00,152.00){\line(-1,-1){7.00}}
\put(75.00,145.00){\line(1,-1){7.00}}
\put(82.00,138.00){\line(-1,-1){7.00}}
\put(75.00,131.00){\line(1,-1){5.00}}
\put(89.00,159.00){\line(-1,-1){7.00}}
\put(82.00,152.00){\line(1,-1){7.00}}
\put(89.00,145.00){\line(-1,-1){7.00}}
\put(82.00,138.00){\line(1,-1){7.00}}
\put(89.00,131.00){\line(-1,-1){5.00}}
\put(89.00,159.00){\line(1,-1){7.00}}
\put(96.00,152.00){\line(-1,-1){7.00}}
\put(89.00,145.00){\line(0,-1){14.00}}
\put(89.00,131.00){\line(1,-1){5.00}}
\put(125.00,158.00){\line(1,-1){7.00}}
\put(132.00,151.00){\line(-1,-1){7.00}}
\put(125.00,144.00){\line(1,-1){7.00}}
\put(132.00,137.00){\line(-1,-1){7.00}}
\put(125.00,130.00){\line(1,-1){7.00}}
\put(133.00,152.00){\line(-1,-1){1.00}}
\put(132.33,151.00){\line(1,-1){6.67}}
\put(139.00,144.33){\line(0,-1){0.33}}
\put(139.00,144.00){\line(-1,-1){7.00}}
\put(132.00,137.00){\line(1,-1){7.00}}
\put(139.00,130.00){\line(-1,-1){7.00}}
\put(61.00,119.00){\line(-1,-1){6.00}}
\put(55.00,113.00){\line(1,-1){12.00}}
\put(67.00,101.00){\line(-1,-1){12.00}}
\put(55.00,89.00){\line(1,-1){6.00}}
\put(61.00,83.00){\line(1,1){6.00}}
\put(56.00,100.00){\line(-1,1){1.00}}
\put(55.00,101.00){\line(1,1){12.00}}
\put(67.00,113.00){\line(-1,1){6.00}}
\put(67.00,113.00){\line(1,-1){6.00}}
\put(73.00,107.00){\line(-1,-1){6.00}}
\put(67.00,101.00){\line(1,-1){6.00}}
\put(73.00,95.00){\line(-1,-1){6.00}}
\put(144.00,111.00){\line(-1,-1){12.00}}
\put(132.00,99.00){\line(1,-1){12.00}}
\put(144.00,87.00){\line(1,1){12.00}}
\put(156.00,99.00){\line(-1,1){12.00}}
\put(138.00,105.00){\line(1,-1){12.00}}
\put(150.00,105.00){\line(-1,-1){12.00}}
\put(61.00,83.00){\line(1,-1){2.00}}
\put(71.00,81.00){\line(0,0){0.00}}
\put(79.00,78.00){\line(-1,-1){6.00}}
\put(73.00,72.00){\line(1,-1){12.00}}
\put(85.00,60.00){\line(-1,-1){12.00}}
\put(73.00,48.00){\line(1,-1){6.00}}
\put(79.00,42.00){\line(1,1){6.00}}
\put(74.00,59.00){\line(-1,1){1.00}}
\put(73.00,60.00){\line(1,1){12.00}}
\put(85.00,72.00){\line(-1,1){6.00}}
\put(85.00,72.00){\line(1,-1){6.00}}
\put(91.00,66.00){\line(-1,-1){6.00}}
\put(85.00,60.00){\line(1,-1){6.00}}
\put(91.00,54.00){\line(-1,-1){6.00}}
\put(89.00,40.00){\line(0,0){0.00}}
\put(91.00,78.00){\line(1,-1){6.00}}
\put(97.00,72.00){\line(-1,-1){6.00}}
\put(91.00,66.00){\line(1,-1){6.00}}
\put(97.00,60.00){\line(-1,-1){6.00}}
\put(91.00,54.00){\line(1,-1){6.00}}
\put(97.00,48.00){\line(-1,-1){6.00}}
\put(70.00,20.00){\line(1,1){9.00}}
\put(79.00,29.00){\line(1,0){18.00}}
\put(97.00,29.00){\line(1,-1){9.00}}
\put(106.00,20.00){\line(-1,-1){9.00}}
\put(97.00,11.00){\line(-1,0){18.00}}
\put(79.00,11.00){\line(-1,1){9.00}}
\put(79.00,29.00){\line(1,-1){18.00}}
\put(79.00,11.00){\line(1,1){18.00}}
\put(44.00,135.00){\makebox(0,0)[cc]{$\rho$}}
\put(43.00,99.00){\makebox(0,0)[cc]{$\rho$}}
\put(43.00,57.00){\makebox(0,0)[cc]{$\rho$}}
\put(43.00,15.00){\makebox(0,0)[cc]{$\rho$}}
\put(124.00,98.00){\makebox(0,0)[cc]{$\rho$}}
\put(3.00,140.00){\makebox(0,0)[cc]{\bf 1a}}
\put(67.00,138.00){\makebox(0,0)[cc]{\bf 1b}}
\put(119.00,136.00){\makebox(0,0)[cc]{\bf 1c}}
\put(3.00,97.00){\makebox(0,0)[cc]{\bf 2a}}
\put(53.00,93.00){\makebox(0,0)[cc]{\bf 2b}}
\put(89.00,95.00){\makebox(0,0)[cc]{\bf 2c}}
\put(136.00,89.00){\makebox(0,0)[cc]{\bf 2d}}
\put(3.00,55.00){\makebox(0,0)[cc]{\bf 3a}}
\put(66.00,51.00){\makebox(0,0)[cc]{\bf 3b}}
\put(3.00,11.00){\makebox(0,0)[cc]{\bf 4a}}
\put(66.00,9.00){\makebox(0,0)[cc]{\bf 4b}}
\put(24.00,153.00){\makebox(0,0)[cc]{$r(\rho)$}}
\put(16.00,116.00){\makebox(0,0)[cc]{$r(\rho)$}}
\put(37.00,119.00){\makebox(0,0)[cc]{$r(\rho)$}}
\put(101.00,115.00){\makebox(0,0)[cc]{$r(\rho)$}}
\put(119.00,88.00){\makebox(0,0)[cc]{$r(\rho)$}}
\put(34.00,77.00){\makebox(0,0)[cc]{$r(\rho)$}}
\put(43.00,141.00){\makebox(0,0)[cc]{$A_1(\rho)$}}
\put(31.00,145.00){\makebox(0,0)[cc]{$A_2(\rho)$}}
\put(44.00,108.00){\makebox(0,0)[cc]{$A(\rho)$}}
\put(45.00,66.00){\makebox(0,0)[cc]{$A(\rho)$}}
\put(46.00,25.00){\makebox(0,0)[cc]{$A(\rho)$}}
\put(75.00,152.00){\makebox(0,0)[cc]{R}}
\put(89.00,152.00){\makebox(0,0)[cc]{R}}
\put(82.00,145.00){\makebox(0,0)[cc]{T}}
\put(77.00,138.00){\makebox(0,0)[cc]{R}}
\put(87.00,138.00){\makebox(0,0)[cc]{R}}
\put(127.00,151.00){\makebox(0,0)[cc]{R}}
\put(132.00,144.00){\makebox(0,0)[cc]{R}}
\put(138.00,99.00){\makebox(0,0)[cc]{R}}
\put(150.00,99.00){\makebox(0,0)[cc]{R}}
\put(144.00,106.00){\makebox(0,0)[cc]{R}}
\put(144.00,93.00){\makebox(0,0)[cc]{R}}
\put(67.00,107.00){\makebox(0,0)[cc]{R}}
\put(61.00,101.00){\makebox(0,0)[cc]{R}}
\put(85.00,66.00){\makebox(0,0)[cc]{T}}
\put(79.00,60.00){\makebox(0,0)[cc]{R}}
\put(91.00,60.00){\makebox(0,0)[cc]{R}}
\put(79.00,20.00){\makebox(0,0)[cc]{R}}
\put(98.00,20.00){\makebox(0,0)[cc]{R}}
\put(88.00,25.00){\makebox(0,0)[cc]{T}}
\put(88.00,15.00){\makebox(0,0)[cc]{T}}
\put(132.00,151.00){\line(1,1){6.00}}
\bezier{68}(21.00,106.00)(16.00,112.00)(7.00,111.00)
\bezier{64}(30.00,106.00)(35.00,112.00)(43.00,111.00)
\put(59.00,81.00){\line(1,1){2.00}}
\put(61.00,83.00){\line(1,-1){2.00}}
\put(67.00,89.00){\line(1,-1){6.00}}
\put(73.00,83.00){\line(-1,-1){2.00}}
\put(67.00,113.00){\line(1,1){6.00}}
\put(73.00,119.00){\line(-1,1){1.00}}
\put(60.00,120.00){\line(1,-1){1.00}}
\put(61.00,119.00){\line(1,1){1.00}}
\put(85.00,72.00){\line(1,1){6.00}}
\put(91.00,78.00){\line(-1,1){1.00}}
\put(78.00,79.00){\line(1,-1){1.00}}
\put(79.00,78.00){\line(1,1){1.00}}
\put(85.00,48.00){\line(1,-1){6.00}}
\put(91.00,42.00){\line(-1,-1){1.00}}
\put(78.00,41.00){\line(1,1){1.00}}
\put(79.00,42.00){\line(1,-1){1.00}}
\put(91.00,78.00){\line(1,1){1.00}}
\put(91.00,42.00){\line(1,-1){1.00}}
\put(55.00,101.00){\line(1,-1){12.00}}
\put(67.00,89.00){\line(0,0){0.00}}
\bezier{200}(10.00,76.00)(25.00,57.00)(43.00,76.00)
\bezier{188}(10.00,35.00)(25.00,18.00)(43.00,35.00)
\put(86.00,106.00){\makebox(0,0)[cc]{$A(\rho)$}}
\put(13.00,35.00){\makebox(0,0)[cc]{$r_1(\rho)$}}
\bezier{164}(8.00,21.00)(25.00,20.00)(44.00,34.00)
\put(9.00,23.00){\makebox(0,0)[cc]{$r_2(\rho)$}}
\put(125.00,159.00){\line(0,-1){36.00}}
\put(73.00,60.00){\line(1,-1){12.00}}
\put(85.00,48.00){\line(0,0){0.00}}
\end{picture}
\caption{\small
  Plots showing the qualitative behaviour of the metric functions and
  Carter-Penrose diagrams for the four different types of regular \ssph\
  \bhs.  Diagrams 1b and 1d (like those for the non-extreme and extreme \RN\
  metrics) refer to curves $A_1$ and $A_2$ in plot 1a, respectively. Diagram
  2b, like that for the extreme Kerr metric, refers to plot 2a, diagram 2d
  to plot 2c, 3b to 3a and 4b to 4a. The R and T letters in the diagrams
  designate the R and T space-time regions. Diagrams 1b, 1c, 2b and 3b are
  infinitely extendible upward and downward. In all diagrams, all inner
  slanting lines depict horizons while all boundaries correspond to
  $r=\infty$, with the following exceptions: the verticals in diagrams 1b
  and 1d describe a regular centre, $r=0$; the horizontals in diagram 4b
  correspond either to $r=\infty$ or to $r = r_0  >0$, according to the
  curves $r_1(\rho)$ or $r_2(\rho)$ at large negative $\rho$.}
\end{figure}

    Type 1 traces back to Bardeen's work \cite{bardeen} which put forward
    the very idea of regular BHs instead of singular ones and suggested, as
    an example, a particular BH configuration with
\beq
    r \equiv \rho, \cm                                      \label{bard}
          A(\rho) = 1 - \frac{M\rho^2}{(\rho^2 + q^2)^{3/2}},
\eeq
    where $M,\ q = \const$, and two horizons exist provided $q^2 < (16/27)
    M^2$. Later on there appeared numerous examples (\cite{dym92,abg,ned01}
    and others) of regular BH solutions where, just as in \eq (\ref{bard}),
    $r \equiv \rho$. This, by virtue of the Einstein equation, implies that
    the stress-energy tensor of the matter source satisfies the condition
    $T^0_0 \equiv T^1_1$, or, in other words, $\eps = -p_r$ ($\eps = T^0_0$
    is the energy density and $p_r = -T^1_1$ is the radial pressure). The
    latter condition is invariant under radial boosts, making it possible to
    ascribe the source to vacuum matter \cite{dym92}, and at a regular
    centre in this case the matter equation of state has necessarily the
    form of a cosmological constant \cite{bdym03}, $T\mN \propto \delta\mN$.

    It was also shown \cite{ned01} that regular BHs with $r\equiv \rho$ and
    any $A(\rho)$ satisfying the regular centre conditions may be obtained
    as magnetic monopole solutions of \GR\ coupled to gauge-invariant
    nonlinear electrodynamics with the Lagrangian $L(F)$, $F := F\mn F\MN$
    ($F\mn$ is the electromagnetic field tensor): the arbitrariness in
    $A(\rho)$ corresponds to the freedom of choosing the function $L(F)$.
    Solutions with an electric charge were shown \cite{ned01} to be
    impossible whatever be the choice of $L(F)$ if $L(F)$ is the same in
    the whole space; this theorem, however, may be circumvented by assuming
    different forms of $L(F)$ near the centre and at large $r$ \cite{bur},
    i.e., by requiring a sort of phase transition(s) at some value(s) of the
    radial coordinate.

    Other examples of type 1 regular BHs have also been found and discussed,
    see, e.g., \cite{reg_bh} and references therein.

    Type 2 regular BHs, with and without an electric charge, have been
    obtained \cite{cold,cold-q} in the framework of the Brans-Dicke
    scalar-tensor theory (STT) with the coupling constant $\omega < -3/2$,
    when the theory is of anomalous, or phantom nature. Their existence
    requires fine tuning in the form of specific relations between $\omega$
    and the integration constants of the corresponding exact solutions. It
    is of interest to note that these regular cold \BD\ BHs have singular
    counterparts in the Einstein frame, in other words, in \GR\ with a
    minimally coupled phantom scalar field (the so-called anti-Fisher
    solution) \cite{pha2}.

    Type 3 regular BHs have been found \cite{casadio,bbh1} as \ssph\
    solutions to the effective equations \cite{SMS99} describing 4D gravity
    in an RS2 type \bw. It has been shown  that such regular solutions are
    generic in a certain range of the integration constants \cite{bbh1} and
    that many of them are stable, at least under some kinds of
    perturbations \cite{abdalla06}. It should be stressed, however, that
    these 4D equations do not form a closed set, to study the full 5D
    geometry of the bulk one should solve the corresponding 5D equations,
    and there are only tentative results in this direction \cite{casa2}.

    Type 4 configurations have been obtained \cite{pha1} as generic
    solutions to the Einstein-scalar equations for the case of minimally
    coupled phantom scalar fields with certain potentials. Such scalar
    fields have recently become popular in the cosmological context
    since they are able to provide an equation of state with the pressure to
    density ratio $p/\eps = w < -1$. It is this kind of equation of state
    that is probably required for the dark energy (DE) component of the
    material content of the Universe to account for its accelerated
    expansion (see, e.g., the reviews \cite{copeland,starob} and references
    therein). Type 4 configurations (we suggest to call them black
    universes) are, in our view, of particular interest, and in what follows
    we will discuss them in some more detail.

\section{Black universes with a minimally coupled scalar field\label{s3}}

    Consider the action for a self-gravitating phantom scalar field in \GR
\beq                                                            \label{act}
     S = \int \sqrt{g}\, d^4 x [R - (\d\phi)^2 - 2V(\phi)],
\eeq
    where $g = |\det(g\mn)|$, $(d\phi)^2 = g\MN\d_\mu\phi \d_\nu\phi$ and
    $V(\phi)$ is an arbitrary potential.  With the metric (\ref{ds}) and
    $\phi=\phi(\rho)$, the scalar field equation and three independent
    combinations of the Einstein equations read
\bear
	 (Ar^2 \phi')' \eql - r^2 dV/d\phi,                  \label{phi}
\\
              (A'r^2)' \eql - 2r^2 V;                          \label{00}
\\
              2 r''/r \eql {\phi'}^2 ;                         \label{01}
\\
         A (r^2)'' - r^2 A'' \eql 2 ,                         \label{02}
\ear
    where the prime denotes $d/d\rho$. The scalar field equation (\ref{phi})
    is a consequence of \eqs (\ref{00})--(\ref{02}), which, given a
    potential $V(\phi)$, form a determined set of equations for the unknowns
    $r(\rho),\ A(\rho),\ \phi(\rho)$. \eq(\ref{02}) can be integrated giving
\bear
          B' \equiv (A/r^2)' = 2(\rho_0 -\rho)/r^4,          \label{B'}
\ear
    where $B(\rho) = A/r^2$ and $\rho_0$ is an integration constant.

    \eq (\ref{B'}) severely restricts the possible dispositions of zeros
    of the function $A(\rho)$ and hence the global causal structure of
    space-time \cite{vac1}.

    Indeed, horizons are regular zeros of $A(\rho)$ and hence $B(\rho)$.
    By (\ref{B'}), $B(\rho)$ increases at $\rho < \rho_0$, has a maximum at
    $\rho=\rho_0$ and decreases at $\rho>\rho_0$. It can have {\it at
    most\/} two simple zeros, bounding a range $B > 0$ (R region), or one
    double zero and two T regions around. It can certainly have a single
    simple zero or no zeros at all.

    So the choice of possible types of global causal structure is precisely
    the same as for the general Schwarzschild-de Sitter solution with
    arbitrary mass and cosmological constant. This result ({\it the Global
    Structure Theorem\/} \cite{vac1}) equally applies to normal and phantom
    fields since \eqs (\ref{02}), (\ref{B'}) are the same for them. It holds
    for any sign and shape of $V(\phi)$ and under any assumptions on the
    asymptotics. BHs with scalar hair (respecting the existing no-hair
    theorems) are not excluded. Examples of (singular) BHs with both normal
    (e.g., \cite{vac2,mann95,salg}) and phantom \cite{lecht} scalar hair are
    known. However, BHs with a regular centre (type 1 according to the
    previous section) are ruled out for our system since their existence
    requires a regular minimum of $B(\rho)$.

    As shown in Ref.\,\cite{pha1}, the system (\ref{phi})--(\ref{02})
    has as many as 16 types of regular solutions with flat, de Sitter and
    AdS asymptotic behaviour. Let us discuss \asflat\ configurations, for
    which $A(\rho) \to 1$ and $r(\rho) \approx \rho$ as $\rho\to\infty$.
    Then, as $\rho$ decreases from infinity, the derivative $r'$ also
    decreases according to \eq (\ref{01}), the decrease rate depending on
    the details of the system. If the decrease is slow enough, $r(\rho)$
    will reach zero at some finite $\rho$, which means that the system has a
    centre. The latter may be regular only if horizons are absent (otherwise
    there would be a type 1 regular BH which is ruled out here), and a
    particlelike solution is then obtained instead of a BH one.

    Assuming the absence of singularities, other opportunities are $r(\rho)
    \to r_0 = \const$ and $r(\rho) \to \infty$ as $\rho \to -\infty$. (Note
    that any kind of oscillatory behaviour of $r(\rho)$ is ruled out by
    (\ref{01}) according to which $r'' \geq 0$.)

    In the first case, according to (\ref{B'}),
\[
     A' \approx - 2\rho/r_0^2 \to +\infty, \cm  A \approx -\rho^2/r_0^2
     \cm {\rm as}  \quad 	\rho  \to -\infty.
\]
    So this ``$r_0$ asymptotic'' is located in a T region. The radius $r_0$
    is related to the limiting value of the potential: $V \to - 1/r_0^2$.
    Changing the notations $-\rho \to T$ and $t\to x$ (since $\rho$ here
    becomes a temporal coordinate and the former time $t$ a spatial one), we
    can write the asymptotic form of the resulting \KS\ metric as
\beq                                                          \label{KS-r0}
      ds^2 \approx \frac{r_0^2}{T^2}\,dT^2 - \frac{T^2}{r_0^2} dx^2
			- r_0^2 d\Omega^2 \cm {\rm as}\quad T\to\infty.
\eeq
    It is a highly anisotropic universe, exhibiting no expansion in the two
    angular directions and an exponential (in terms of the physical time
    $\tau \sim \log T$) expansion in the third direction $x$. From the
    viewpoint of an observer at large positive $\rho$, this universe is
    located beyond the event horizon of a BH.

    In case $r\to \infty$, the most interesting opportunity is that
    $r \sim |\rho|$ as $\rho\to -\infty$. Assuming $r \approx -a\rho$,
    $a =\const >0$, from (\ref{B'}) and (\ref{00}) we obtain
\beq
      A \approx  1/a^2 - C a^2 \rho^2, \cm
                   V \approx  3 C a^2, \cm C = \const.
\eeq
    It is easy to verify that $C=0$ leads to a Minkowski metric at large
    negative $\rho$ (though the time rate will be different from that at
    large positive $\rho$ if $a\ne 1$), an anti-de Sitter metric if $C < 0$
    and a de Sitter metric if $C > 0$. In the cases $C \leq 0$ horizons are
    absent since otherwise the function $B(\rho)$ would have a minimum at
    some finite $\rho$, which cannot happen due to (\ref{B'}). Thus
    possible solutions with $C\leq 0$ describe traversable \whs, and the
    details of their geometry depend on the particular choice of $V(\phi)$.

    Lastly, for $C > 0$ we obtain a de Sitter asymptotic behaviour of the
    solution, describing (since we are now in a T region) isotropic
    expansion or contraction. From the viewpoint of an external observer,
    located at large positive $\rho$, it is a BH, but a possible BH explorer
    now has a chance to survive for a new life in an expanding, gradually
    isotropizing \KS\ universe. The specific BH profile and the
    isotropization regime after crossing the horizon depend on the choice of
    $V(\phi)$.

    Since, according to the Global Structure Theorem (see above), an
    \asflat\ configuration can have only one simple horizon, such a BH has a
    Schwarzschild-like causal structure, but the singularity $r=0$ in the
    Carter-Penrose diagram is now replaced by the de Sitter infinity
    $r=\infty$.

    A simple example may be obtained by putting \cite{pha1}
\beq                                                         \label{r1}
	r = (\rho^2 + b^2)^{1/2}, \cm b = \const > 0.
\eeq
    and using the inverse problem scheme. \eq (\ref{B'}) gives
\bearr  \nq                                                   \label{B1}
      B(\rho) = \frac{A(\rho)}{r^2(\rho)}
      = \frac{c}{b^2} + \frac{1}{b^2+\rho^2} + \frac{\rho_0}{b^3}
	\left(\frac{b\rho}{b^2 + \rho^2} + \tan^{-1}\frac{\rho}{b}\right),
\ear
    where $c = \const$. \eqs (\ref{01}) and (\ref{00}) then lead to
    expressions for $\phi(\rho)$ and $V(\rho)$:
\bear                                                        \label{phi1}
      \phi \eql \pm\sqrt{2} \tan^{-1} (\rho/b) + \phi_0,
\\  \nhq                                                         \label{V1}
     V \eql - \frac{c}{b^2} \frac{r^2 + 2\rho^2}{r^2}
	   - \frac{\rho_0}{b^3}
	   	\left( \frac{3b\rho}{r^2}
    + \frac{r^2 + 2\rho^2}{r^2}\tan^{-1}\frac{\rho}{b}\right)\quad\
\ear
    with $r = r(\rho)$ given by (\ref{r1}). In particular,
\beq                                                       \label{BV_as}
      B(\pm \infty) = -\frac 13 V(\pm \infty)
			= \frac{2bc \pm \pi \rho_0}{2b^3}.
\eeq
    Choosing in (\ref{phi1}), without loss of generality, the plus sign and
    $\phi_0=0$, we obtain for $V(\phi)$ ($\psi := \phi/\sqrt{2}$):
\bearr
     V(\phi) = -\frac{c}{b^2} (3 - 2\cos^2 \psi)             \label{Vf}
     - \frac{\rho_0}{b^3} \left[3\sin\psi \cos\psi
         			+ \psi (3 - 2\cos^2 \psi) \right].
\ear

    The solution behaviour is controlled by two integration constants: $c$
    that moves $B(\rho)$ up and down, and $\rho_0$ showing the maximum of
    $B(\rho)$. Both $r(\rho)$ and $B(\rho)$ are even functions if
    $\rho_0=0$, otherwise $B(\rho)$ loses this symmetry. Asymptotic flatness
    at $\rho = +\infty$ implies $2bc = -\pi \rho_0$ while the Schwarzschild
    mass, defined in the usual way, is $m =\rho_0/3$.

    Under this asymptotic flatness assumption, for $\rho_0 = m =0$ we obtain
    the simplest symmetric configuration, the
    Ellis \wh\ \cite{h_ell}: $A \equiv 1$, $V \equiv 0$. For $\rho_0 < 0$,
    according to (\ref{BV_as}), we obtain a \wh\ with $m < 0$ and an AdS
    metric at the far end, corresponding to the cosmological constant $V_- <
    0$. For $\rho_0 > 0$, when $V_{-} > 0$, there is a regular BH with $m >
    0$ and a de Sitter asymptotic far beyond the horizon, precisely
    corresponding to the above description of a black universe.

    The horizon radius $r(\rho_h)$ may be obtained by solving the
    transcendental equation $A(\rho_h) =0$, where $A(\rho)$ is given
    by \eq (\ref{B1}). It depends on both parameters $m$ and $b = \min
    r(\rho)$ and cannot be smaller than $b$, which also plays the role of a
    scalar charge:  $\psi \approx \pi/2 -b/\rho$ at large $\rho$.  Since
    $A(0) = 1 + c$, the throat $\rho=0$ is located in the R region if $c
    >-1$, i.e., if $3\pi m < 2b$, at the horizon if $3\pi m = 2b$ and in the
    T region beyond it if $3\pi m > 2b$. The above relations between $m$ and
    $b$ show (and it is probably generically true) that if the BH mass
    dominates over the scalar charge, the throat is invisible to a distant
    observer, and the BH looks from the static region almost as usual in
    general relativity.

    As follows from \eqs (\ref{phi}) and (\ref{00}), the potential $V$
    tends to a constant and, moreover, $dV/d\phi \to 0$ at each end of the
    $\rho$ range. It is a general property of all classes of regular
    solutions indicated in \cite{pha1}. More precisely, {\it a regular
    scalar field configuration requires a potential with at least two
    zero-slope points (not necessarily extrema) at different values of
    $\phi$}.

    Suitable potentials are, e.g., $V = V_0 \cos^2 (\phi/\phi_0)$
    and the Mexican hat potential $V = (\lambda/4)(\phi^2 - \eta^2)^2$ where
    $V_0,\ \phi_0,\ \lambda,\ \eta$ are constants. A flat infinity at $\rho
    = +\infty$ certainly requires $V_+ = 0$, while a de Sitter asymptotic
    can correspond to a maximum of $V$ since phantom fields tend to climbing
    up the slope of the potential rather than rolling down, as is evident
    from \eq (\ref{phi}).  Accordingly, Faraoni \cite{fara05}, considering
    spatially flat isotropic phantom cosmologies, has shown that if
    $V(\phi)$ is bounded above by $V_0 = \const > 0$, the de Sitter solution
    is a global attractor. Very probably this conclusion extends to \KS\
    cosmologies after isotropization.

    One more point may be stressed: the late-time de Sitter expansion rate
    is entirely determined by the corresponding potential value $V_- >0$
    (which, in our notation according to (\ref{act}), coincides with the
    effective cosmological constant $\Lambda$ at late times) rather than
    by the details of the solution such as the Schwarzschild mass defined at
    the flat asymptotic.

    We can conclude that black universes are a generic kind of solutions to
    the Einstein-scalar equations in the case of phantom scalars with proper
    potentials.

\section {Black universes with other forms of phantom matter\label{s4}}

    Besided the minimally coupled scalar field (\ref{act}), there are a
    number of other suggestions for modelling the possible phantom
    behaviour of dark energy, see, e.g., the review \cite{copeland} and
    references therein.

    Let us show that type 4 regular BHs, or black universes, are, under
    appropriate additional conditions, solutions to the field equations of
    at least two large classes of such theories: scalar-tensor theories
    (STT) of gravity, or theories with nonminimally coupled scalar fields,
    and the class of models called k-essence. We will show that
    black-universe solutions may be found in both these classes.

\subsection{Scalar-tensor theories}

     Consider a general (Bergmann-Wagoner-Nordtvedt) STT, defined in a
     space-time manifold with the metric $g\mn$ (called the Jordan conformal
     frame), for which the gravitational action is written of the form
\beq
     S_{\rm STT} = \int d^4 x \sqrt{g}                    \label{act-J}
                  [f(\Phi) R + h(\Phi) (\d\Phi)^2 -2U(\Phi)],
\eeq
     where $g =|\det(g\mn)|$, $(d\Phi)^2 = g\MN\, \d_{\mu}\Phi \d_{\nu}\Phi$
     and $f,\ h,\ U$ are arbitrary functions of the scalar field $\Phi$.

     The action (\ref{act-J}) is simplified by the well-known
     conformal mapping \cite{wagon}
\bear
     g\mn \eql \og\mn/|f(\Phi)|,                          \label{g-wag}
\\
     \frac{d\phi}{d\Phi} \eql \pm
                \frac{\sqrt{|l(\Phi)}|}{f(\Phi)},         \label{Phi-wag}
\cm
     l(\Phi) := fh + \frac{3}{2}\Bigl(\frac{df}{d\Phi}\Bigr)^2,
\ear
     removing the nonminimal scalar-tensor coupling express\-ed in the
     factor $f(\Phi)$ before $R$. The action (\ref{act-J}) is now
     specified in a new manifold with the metric $\og\mn$
     (the Einstein frame) and the new scalar field $\phi$:
\bearr
     S_{\rm E} = \int  d^4 x \sqrt{\og}                     \label{act-E}
        \Bigl\{\sign f \bigl[\oR
                + (\sign l) (\d\phi)^2\bigr] - 2V(\phi)\Bigr\},
\ear
     where the determinant $\og$, the scalar curvature $\oR$ and
     $(\d\phi)^2$ are calculated using $\og\mn$, and
\beq
        V(\phi) = |f(\Phi)|^{-2}\, U(\Phi).                  \label{UV}
\eeq
     The action (\ref{act-E}) is similar to that of GR with a minimally
     coupled scalar field $\phi$ but contains two sign factors. The usual
     sign of gravitational coupling corresponds to $f > 0$, while $l(\Phi)$
     distinguishes normal ($l >0$) and phantom ($l < 0$) fields. Of interest
     for us are phantom field theories with usual gravitational coupling,
     i.e., we will suppose $f > 0$, $l < 0$.

     Then the vacuum field equations have precisely the form
     (\ref{phi})--(\ref{02}), with all the corresponding solutions. The main
     properties of these solutions are preserved after transformation back
     to the Jordan frame, provided the conformal factor $1/f$ is everywhere
     nonzero and regular. Indeed, in this case, a flat asymptotic transforms
     to a flat asymptotic (although maybe with a different Schwarzschild
     mass) and a horizon to a horizon of the same order. Thus one can
     guarantee that a black universe solution in the Einstein frame
     preserves its qualitative properties in the Jordan frame. However,
     some remarks are in order.

     First, the asymptotic regimes in the Einstein frame occur at values of
     $\phi$ where $V(\phi)$ has zero slope and, moreover, asymptotic
     flatness requires $V = 0$. Assuming $f \ne 0$, we may reformulate these
     requirements in terms of $U(\Phi)$: evidently, $U = 0$ at the flat
     asymptotic in the Jordan frame, and, at the other end, it is the
     function $V(\phi)$ given by \eq (\ref{UV}) that must have zero slope,
     while nothing certain may be said about $U$ at the corresponding value
     of $\Phi$.

     Second, a de Sitter asymptotic behaviour at $\rho \to -\infty$ is not
     necessarily preserved in Jordan's frame even if it takes place in
     Einstein's. In general, the expansion law will be different but
     izotropization must take place. Everything depends on the specific
     properties of $U$ and $f$.

     In general, we see that black-universe solutions may be expected in a
     generic STT with a phantom scalar field provided the potential function
     $U(\phi)$ (or, as it is sometimes said, the scalar-dependent
     ``cosmological constant'') satisfies some natural conditions which are
     best formulated in terms of the Einstein conformal frame.

\subsection {k-essence}

     Another large class of scalar field ($\phi$) models invoked in order to
     describe the modern cosmological acceleration is characterized by a
     non-canonical dependence of the action on the derivatives $\d_\mu\phi$.
     The most general form of such actions for scalar fields minimally
     coupled to space-time curvature may be written as
\beq                                                        \label{act-k}
       S = \int \d^4 x \sqrt{g} [R + F(\phi, X)], \cm
       X:= (\d\phi)^2
\eeq
     (see, e.g., \cite{copeland}; references to numerous papers discussing
     particular cases of k-essence as well as other models of dark energy
     can also be found in this comprehensive review).

     The stress-energy tensor of $\phi$ then has the form
\beq                                                         \label{SET-k}
       T\mN [\phi] = F_X\, \d_\mu\phi \d^\nu\phi - \half F \delta\mN,
\eeq
     where $F_X \equiv \d F/\d X$. Obvious special cases of (\ref{act-k})
     are
\begin{description}
\item[(i)]
        normal and phantom scalar fields with the usual kinetic term
	($F = \ep X - 2V(\phi)$, $\ep = 1$ for a normal field and $\ep = -1$
	for a phantom field);
\item[(ii)]
	the most frequently used forms of k-essence, with
	$F = V(\phi) U(X)$, and, in particular, the so-called tachyonic
	field, for which $F = -V(\phi) \sqrt{1 - X}$.
\end{description}
      In the cosmological setting, the action (\ref{act-k}) may also
      describe such forms of dark energy as
\begin{description}
\item[(iii)]
      	a perfect fluid with the barotropic equation of state $w = \eps/p
	=\const < -1/3$, for which one should put
   \beq                                                    \label{flu}
	  F(\phi, X) = V(\phi) X^{(w+1)/(2w)},
   \eeq
	with an arbitrary function $V(\phi)$;
\item[(iv)]
	a generalized Chaplygin gas, defined as a perfect fluid with the
	equation of state $p = -A/\eps^\alpha$, with $A,\alpha = \const$,
	for which one should choose $F(\phi,X)$ satisfying the equation
   \beq
	  2X \,F_X = F + [-F/(2A)]^{-1/\alpha}.
   \eeq
	In particular, to obtain a ``simple'' Chaplygin gas, with the
	equation of state $p = -A/\eps$, one can put
   \beq
	  F = \pm 2\sqrt{A}\,\sqrt{1 + Xf(\phi)}          \label{chap}
   \eeq
	with an arbitrary function $f(\phi)$.
\end{description}
    To verify this, it is sufficient to put $\phi = \phi(t)$ and $g_{00}=1$,
    where $x^0 = t$ is the cosmological time, so that $X = (d\phi/dt)^2$; we
    have then $\eps = XF_X - F/2$ and $p = F/2$.

    Let us now return to \ssph\ configurations with the metric (\ref{ds}).
    It is then straightforward to check that the field equations for the
    action (\ref{act-k}) may be written in the following form similar to
    (\ref{phi})--(\ref{02}):
\bear
	 2(F_X A r^2 \phi')' \eql -r^2 F_\phi;                \label{phi-k}
\\
              (A'r^2)' \eql r^2(F - XF_X);                    \label{00-k}
\\
              2 r''/r \eql -{\phi'}^2\, F_X,                   \label{01-k}
\\
         A (r^2)'' - r^2 A'' \eql 2.                           \label{02-k}
\ear
    Again the scalar field equation (\ref{phi-k}) may be obtained from the
    Einstein equations (\ref{00-k})--(\ref{02-k}), and so the latter three
    may be considered as a determined set of equations for $\phi(\rho)$,
    $A(\rho)$ and $r(\rho)$ if the function $F(\phi,X)$ has been prescribed.

    \eq (\ref{02-k}) is the same as (\ref{02}) due to the relation
    $T_0^0 = T^2_2$ valid for the tensor (\ref{SET-k}) just as for that of a
    usual minimally coupled scalar. Therefore, we again obtain the integral
    (\ref{B'}) and, as its consequence, the Global Structure Theorem
    \cite{vac1} restricting the possible global properties of the solutions.

    Since the action (\ref{act-k}) is more general than (\ref{act}), it is
    reasonable to expect that the set of possible solutions will also be
    richer. Let us try to formulate some necessary conditions for obtaining
    solutions of interest for us here, namely, black universes.

    Above all, to obtain $r\to\infty$ as $\rho\to \pm \infty$ it is
    necessary to have $r'' > 0$ at least in some range of $\rho$, hence
    $F_X < 0$ (the ``phantom condition'').

    Other conditions may be deduced by assuming $r \sim \rho$ as $\rho\to
    \pm \infty$ and requiring a Minkowski limit at large positive $\rho$ and
    a de Sitter limit with some cosmological constant $\Lambda_-$ at large
    negative $\rho$. Let us also assume that $F$ is a smooth function of
    both arguments while $\phi$ tends to finite limits $\phi_{\pm}$ and
    admits expansions in power series in $1/\rho$ as $\rho\to \pm \infty$.

    Substitution into the field equations then gives:
\bearr                                                     \label{cond-F}
	F,\ F_{\phi},\ XF_X = O(\rho^{-4})\cm {\rm as}\quad \rho\to\infty,
\nnn
     F \to \sqrt{\Lambda_-/3}, \cm  XF_X,\ F_\phi = O(\rho^{-2})
	\cm {\rm as}\quad  \rho\to -\infty.
\ear
    The symbol $O(x)$ means here a quantity of order $x$ or smaller.
    In all cases $X\to 0$ at large $|\rho|$, and so the function $F(\phi, 0)$
    has zero slopes at $\phi = \phi_{\pm}$, just as the potential $V (\phi)$
    of a minimally coupled scalar field in \sect \ref{s3}.

    The asymptotic regimes of $F(\phi,X)$ are different in Minkowski and de
    Sitter limits. Indeed, if we assume $\phi = \phi_+ + \const/\rho$ and
    $\phi' \sim \rho^{-2}$ as $\rho\to \infty$, which is characteristic of a
    massless (long-range) scalar field, then (\ref{cond-F}) leads to
    $F = O(\rho^{-4})$, so that $F(\phi, 0) \sim (\phi - \phi_+)^4$ or is
    even smaller. This is not surprising since $F(\phi, 0)$ behaves as a
    potential which must be smaller than quadratic for a massless behaviour
    of $\phi$.

    On the other hand, at the de Sitter asymptotic, under similar
    assumptions ($\phi = \phi_- +\const/\rho$ as $\rho \to -\infty$), we
    obtain a generic quadratic behaviour of $F$:
    $F(\phi,0) \sim (\phi-\phi_-)^2$, and the massless case is not
    distinguished in the behaviour of the ``potential'' $F$.

    These observations are confirmed by the asymptotic behaviour of the
    potential (\ref{Vf}) in the above example (\ref{r1})--(\ref{Vf}):
    $V \sim (\phi - \pi/\sqrt{2})^4$ as $\rho \to \infty$ and
    $V - V_- \sim (\phi + \pi/\sqrt{2})^2$ as $\rho \to -\infty$ where
    $V_- = \Lambda_- = V\big|_{\rho\to -\infty} = 3\pi\rho_0/b^3$.

    The conditions (\ref{cond-F}) are easily re-formulated for the
    frequently used kinds of k-essence: $F = f(X) - 2V(\phi)$ and
    $F = f(X) V(\phi)$.

    It, however, can be easily shown that black-universe solutions are
    absent for the perfect fluid representations (\ref{flu}) and
    (\ref{chap}). Indeed, $X = -A \phi'{}^2$ changes its sign at the
    horizon, therefore $F$ must make sense for any sign of $X$. Meanwhile,
    in the function (\ref{flu}), the exponent $(w+1)/(2w)$ is fractional
    for any $w < -1$, and the whole expression loses its meaning at $X < 0$.
    As to the function (\ref{chap}), it is nonzero at $X = 0$, which is
    incompatible with asymptotic flatness.

    Thus many, though not all, kinds of k-essence lead to black-universe
    solutions, under necessary conditions similar to those formulated for
    minimally coupled scalar fields.

\section{Conclusion\label{s5}}

    In this paper, we have classified the possible geometries of \ssph,
    \asflat\ BHs and discussed in more detail their particular type, the
    black universes. Such hypothetical configurations combine the properties
    of a wormhole (absence of a centre, a regular minimum of the area
    function) and a \bh\ (a Killing horizon separating R and T regions).

    Quite evidently, such unusual objects require unusual matter for their
    existence. It turns out that they are naturally obtained if one
    considers \ssph\ distributions of phantom matter, which is a subject
    of vast discussion in modern cosmology. We have considered a
    sufficiently general frameworks for the description of phantom matter,
    namely, STT of gravity and the so-called k-essence, and found necessary
    conditions for the existence of black-universe solutions. A conclusion
    is that a great number of reasonable phantom matter models produce such
    solutions.

    The latter lead to the idea that our Universe could appear from
    phantom-dominated collapse in another, ``mother'' universe and undergo
    isotropization (e.g., due to particle creation) soon after crossing the
    horizon. It is known that a \KS\ nature of our Universe is not
    excluded observationally \cite{craw} if its isotropization had happened
    early enough, before the last scattering epoch (at redshifts $z\gtrsim
    1000$). One can notice that we are thus facing one more mechanism of
    universes multiplication, in addition to the well-known mechanism
    existing in the chaotic inflation scenario.

    Somewhat similar ideas on possible appearance of baby universes inside
    BHs have already been discussed from different standpoints
    \cite{tsey92,univ-bh} (see also references therein), and some
    examples of two-dimensional regular BH metrics with properties
    more or less similar to ours are known \cite{tsey92,stoj98,eas03},
    It can be remarked, however, that in two dimensions, where space is
    simply a line, the notion of a centre,
    which plays an essential role in the four-dimensional picture, cannot
    be properly introduced. So it seems that the black-universe solutions
    considered here and in Ref.\,\cite{pha1} are qualitatively new and (if
    certainly amended by adding realistic matter ingredients) may even
    lead to viable alternatives to the existing cosmological scenarios.

\subsection*{Acknowledgments}

	The authors acknowledge partial financial support from DFG project
        436/RUS 113/807/0-1(R) and also (KB and VM) from RFBR Grant
	05-0217478n. KB and VM are thankful to the colleagues from the
	University of Konstanz for kind hospitality.

\end{document}